\documentclass{sig-alternate}
\usepackage{epsfig}
\usepackage{times}
\usepackage[table,xcdraw]{xcolor}
\usepackage{graphicx}
\usepackage{url}
\usepackage{listings}
\usepackage{comment}
\usepackage{graphicx}
\usepackage[english]{babel}
\usepackage{algorithm}
\usepackage[noend]{algpseudocode}
\usepackage{url}
\usepackage{amssymb}
\usepackage{float}
\usepackage{verbatim}

\newcommand{\ie}{i.e.} 
\newcommand{\eg}{e.g.} 
\newcommand{\et}{et al. }
\begin{document}

\conferenceinfo{}{}
\CopyrightYear{2014} 
\crdata{978-1-4503-2627-8/14/07\$15.00.\\
http://dx.doi.org/10.1145/xxxx.xxxx} 

\title{Two-level Data Staging ETL for Transaction Data}

\numberofauthors{1}
\author{
\alignauthor
Xiufeng Liu\\
       \affaddr{University of Waterloo, Canada}\\
       \email{xiufeng.liu@uwaterloo.ca}
}

\maketitle
\begin{abstract}
In data warehousing, Extract-Transform-Load (ETL) extracts the data from data sources into a central data warehouse regularly for the support of business decision-makings. The data from transaction processing systems are featured with the high frequent changes of insertion, update, and deletion. It is challenging for ETL to  propagate the changes to the data warehouse, and maintain the change history. Moreover, ETL jobs typically run in a sequential order when processing the data with dependencies, which is not optimal, \eg, when processing early-arriving data. In this paper, we propose a two-level data staging ETL  for handling transaction data. The proposed method detects the changes of the data from transactional processing systems, identifies the corresponding operation codes for the changes, and uses two staging databases to facilitate the data processing  in an ETL process. The proposed ETL provides the ``one-stop" method for fast-changing, slowly-changing and early-arriving data processing. 
\end{abstract}



\section{Introduction}
\label{sec:intro}
A data warehouse (DW)  is the decision-making database which holds the data extracted from transaction systems, operational data stores and external sources.  The data are processed by the ETL from the source systems into the DW,  traditionally referred as to data warehousing \cite{kimball2004}. The transformed data in the DW \cite{inmon2002} are usually structured according to star schemas and accessed by decision support systems, such as Online Analytical Processing (OLAP) tools and Business Intelligence (BI) applications \cite{march}. 

Data warehousing systems run ETL jobs at a regular interval, such as daily, weekly or monthly. The data source systems such as  transnational databases typically deal with the dynamic data, which means that the data are changed frequently, such as an online shopping system. The data in the  transaction systems are changed during everyday's business operations, \eg, it can be adding a new order, or updating records as simple as an address change of a registered member, or more complex purchase data. The data warehouse of keeping the detailed transaction history is called the {\em system of record} (SOR), which can date back the changes of the data in the transaction processing systems \cite{inmon2003}. For example, business users can retrieve each of the orders, as well as its status, including ordering, delivery, payment, and after-sale services; or cancellation of orders. The changes in the transaction processing systems are reflected in the SOR data through the ETL processes, which involve the operations such as new records addition, update, deletion,  lookup, and loading. The ETL process that supports the high frequent changing data is not trivial, and worth the further study.
\begin{figure*}[htp]
\centering
\epsfig{file=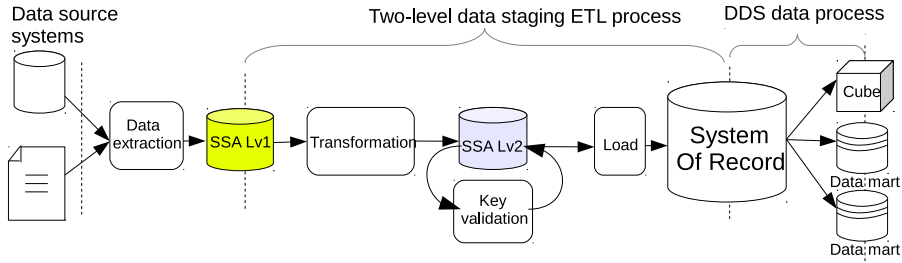, scale=0.75}
\vspace{-10pt}
\caption{The overview of two-level data staging ETL}
\label{fig:architecture}
\vspace{-10pt}
\end{figure*}

In addition, when an ETL process runs, the data fed to the ETL process usually are required to follow a certain order. A typical example is that dimension processing is prior to the fact processing due to the foreign-key relationship between fact and dimension tables. A fact record in the fact table consists of dimension key values and measures (or the facts). When the ETL process processes a fact record, it gets the dimension key values by {\em looking up} the dimension tables where all the dimension values have been loaded. If the fact records arrive at the data warehouse earlier than the dimension records, the {\em lookup} operation will fail. To ensure the integrity of early-arriving data, the parent tables (referenced tables) are loaded first, followed by the child tables (the tables with the foreign keys). Using this approach, however, the downside is obvious. For example,  we have to use some extra space to keep the early-arriving data temporarily.  The waiting wastes the time, and might mess up with the overall schedule plan in a busy IT platform.

In this paper, we use two-sequential staging databases, in combination with different operation codes, to process transaction data. Specifically, this paper makes the following contributions: 1) We propose the two-level data staging ETL method for handling the high frequent changes of data, \ie, with insertion, update and deletion; 2) We propose the method of using different operation codes to support ETL actions, \ie, the method can identify  data changes into different operation codes, and based on the codes for doing the foreign-key validation, transformation, and loading. 3) We propose the argumentation process for handling early-arriving data; and ``one-stop" method for fast- and slowly-changing data processing.

This paper is structured as follows. Section~\ref{sec:overview} gives the overview of the two-level data staging ETL method. Section~\ref{sec:dwdataprocess} details the proposed method, including data change detection, transformation, key validation and loading. Section~\ref{sec:relatedwork} presents the related work, and the state-of-the-art of data warehousing technologies. Section~\ref{sec:conclusion} concludes the paper.

\section{Overview}
\label{sec:overview}
We now describe the two-level data staging ETL for handling  transaction data (see  Figure~\ref{fig:architecture}). The ETL process makes use of two databases for data staging, and completes the necessary data transformation upon the two staging databases. The whole process consists of change detection, transformation, key validation and loading. The ETL process first extracts the data from  source systems (\eg, a transaction processing system) into the level-1  staging database (SSA Lv1), then transposes the data from SSL Lv1 to the level-2 staging database (SSA Lv2). SSA Lv1 maintains a similar database schema as the source system, while  SSA Lv2 maintains a similar schema as the data warehouse (\ie, SOR). In the change detection phase, the data changes that occurred in the transaction processing systems are identified into different operation codes for guiding the ETL actions of data transformation and loading. In the key validation phase, the transposed data in the SSA Lv2 tables are validated based on the foreign-key relationship between tables. During the loading phase, the data in SSA Lv2 tables are well-prepared, and loaded into the SOR tables. The SOR keeps the finest granular records, and the change history of the data. The data in SOR are used to create data marts or cubes through the dimensional data store (DDS) process of aggregation. In the following sections, we focus on the discussion of the two-level data staging ETL process of extracting transaction data from the source systems to the SOR, and briefly introduce the DDS data process.

SSA Lv1 is  used because it is better to extract all the required source data to a staging area. In case the ETL process needs to be re-run, it can still base on the data in SSA Lv1 which was previously extracted from the source tables. SSA Lv2 is used to simplify the ETL process for handling the changes of the data, which might include  fast-changing, slowly-changing,  early-arriving, and static data. As such, the ETL process could simply move the data from SSA Lv2 to SOR without needing any further data transformation, which greatly simplifies the loading process.

\begin{figure*}[htp]
\centering
\epsfig{file=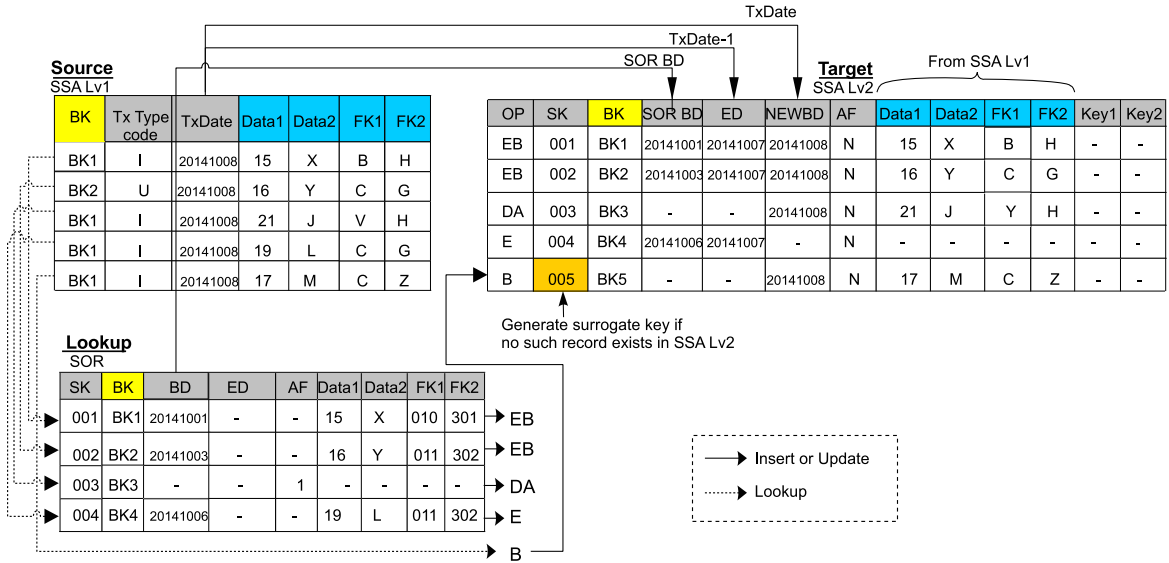, scale=1}
\vspace{-10pt}
\caption{The transformation}
\label{fig:transformation}
\vspace{-10pt}
\end{figure*}

\section{Two-level Data Staging ETL}
\label{sec:dwdataprocess}
We now describe how to process the records from an operational source system into SOR using the two-level data staging ETL.

\subsection{Change Detection}
\begin{figure}[htp]
\centering
\vspace{-10pt}
\epsfig{file=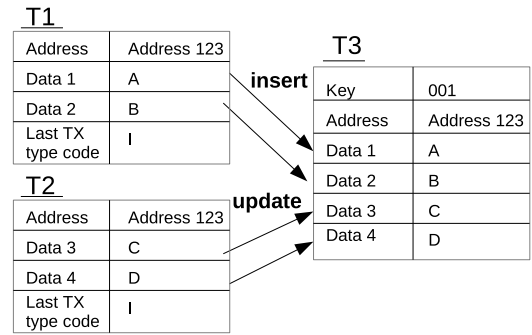, scale=0.8}
\vspace{-10pt}
\caption{Many-to-one mapping}
\label{fig:changedetect}
\vspace{-10pt}
\end{figure}

 A record from the source system is extracted to the SSA Lv1, and marked with the transaction (Tx) type code, ``I", ``U" or ``D", which indicates the action of insertion, update, or deletion in the source system. However, we cannot directly insert or update the record into a SOR table based on the Tx type code. Instead, we need to detect the change of the data (or the Tx type) in the source system, and apply {\em a new operation code} for the change. The reasons are as follows. First, the transaction type code only represents the type of the action happened in the source table, it does not fully apply to the detailed records in the SOR. For example, for the many-to-one mapping as shown in Figure~\ref{fig:changedetect}, in the source system the records in the tables, $T1$ and $T2$, with the transaction type code ``I" are mapped to a single record in $T3$ under the same business key value "Address 123". If the record from $T1$ is first inserted into the target table $T3$. Then, if we solely follow the last Tx type code in $T2$ to insert the record, we will receive an error since it should be ``update", instead of ``insert". The second reason is that the daily ETL job will only extract the latest action applied to the source table. If there is more than one action applied to the source table at the same day, the extraction will miss all the previous actions. For example, if the Tx type codes, ``I" and ``U", are applied to the source table at the same date (\ie, a record is inserted, and updated subsequently), when the ETL  job runs at the end of the day, and updates the SOR table based on the last Tx type code ``U", errors may occur because it should do the insertion, instead of updating the record.
\begin{table}[htp]
\vspace{-10pt}
\centering
  \caption{The operation codes}
  \begin{tabular}{ |p{1.5cm} | p{6cm}| }  \hline
  \rowcolor[HTML]{C0C0C0} 
   {\bf Code}&{\bf Description}  \\ \hline
     {\bf B} & Begin Operation  \\ \hline
     {\bf EB} & End Begin Operation\\ \hline
     {\bf E} & End Operation \\ \hline 
     {\bf A}& Argument Operation\\ \hline 
     {\bf DA}& Deactivate Argument Operation\\ \hline 
  \end{tabular}
  \label{tab:opcodes}
  \vspace{-5pt}
\end{table}

Therefore, we first need to determine whether the record exists in SOR or not, even if the Tx type code is ``I" or ``U" in the source system.  To simplify  handling the changes, we make use of the second level staging database, \ie, SSA Lv2, and detect the correct changes  before loading the record into SOR. We, therefore, move the record from SSA Lv1 to SSA Lv2, and assign a new operation code to the record. The possible operation codes are listed  in Table~\ref{tab:opcodes}. Begin (B) is used to label the record that is newly added into the SOR with a new assigned surrogate key.  End and Begin (EB) is used to label the record for update. It also represents that there exists a previous version of the record in the SOR (with the same business key value), which should be marked as the ``history" by setting a new end date. Then, a new record with the updated values, but  with the same surrogate key as the history record is created.  End (E) is used to label the record that should be marked as ``history" in the SOR by setting the end date (no new record is created). Argument (A) is used to mark an argument record for handling early-arriving data (or the dummy record inserted into the SOR table without values in all attributes except the surrogate and business key). And,  Deactivate Augment (DA) is used to label the record that was previously created in the augmentation. The augment flag in the record will be turned off and  the blank attributes will be updated.

\subsection{Transformation}
\label{sec:transformation}
We have detected the changes of the data, and identified the corresponding operation codes  that will be applied to the SOR. We now detail the transformation process of labeling the records with the operation codes (see Algorithm~\ref{alg:transformation}). For each record from SSA Lv1, the ETL process first looks up the SOR table to check its existence using the business key (BK) (see line~2). Based on the result of the lookup and the transaction type code, we then decide which operation code should be applied (see line~4--11 and 13--19). However, it is possible that the record has already been labelled, and added into SSA Lv2. Thus, we should update the record in SSA Lv2, instead of inserting a new one. This can be done by the following two approaches. The first approach is to look up the SSA Lv2 table, then decide either insert or update action should be done on it. But this involves further lookup logic. The second approach is to ``merge" the two records (some ETL tools call it as ``upsert"). This operator is available in many ETL tools, such as Informatica and Pentaho PDI. 

During the transformation process, business entities are allocated a surrogate key to replace the natural primary key, which is simply to assign a sequence number to the record if it is to be newly created in the SOR.  If the extracted record has the last record transaction type code ``I" or ``U", and cannot be found in the SOR table by the lookup using the business key, it will be identified as a new record. However, we also need to look up the SSA Lv2 table to check its existence. If the record does not exist, we assign a surrogate key to the record, and insert it into the SSA Lv2 table (see line~14--19).

\begin{algorithm}[htp]
\caption{Change detection and transformation}
\label{alg:transformation}
{\scriptsize
\begin{algorithmic}[1]
  \Require{Data has been loaded into SSA Lv1}
  \State $row \gets$ Read a record from SSA Lv1
  \State $found$ $\gets$ Lookup SOR table by $row[BK]$
  \If{$found$}
  	\If{$row[TxCode]=D$}
  		\State Transpose and merge SSA Lv2 with $row[OP]=E$	
  	\Else \Comment{$row[TxCode]=U$ or $I$ }		
  		\State $arg \gets$ Is an argument record in SOR?
  		\If{arg}
  			\State Transpose and merge SSA Lv2 with $row[OP]=DA$
  		\Else
  			\State Transpose and merge SSA Lv2 with $row[OP]=EB$
  		\EndIf
  	\EndIf
  \Else
  	\If{$row[TxCode]=U$ or $I$}
  		\State $found \gets$ Look up SSA Lv2 by $row[BK]$
  		\If{found}
  			\State Transpose and merge SSA Lv2 with $row[OP]=B$
  		\Else
  			\State $row[SK] \gets$ Get a new surrogate key
  			\State Transpose and insert SSA Lv2 with $row[OP]=B$
  		\EndIf
  	\EndIf
  \EndIf
\end{algorithmic}
}
\vspace{-2pt}
\end{algorithm}

\begin{figure*}[htp]
\vspace{-5pt}
\centering
\epsfig{file=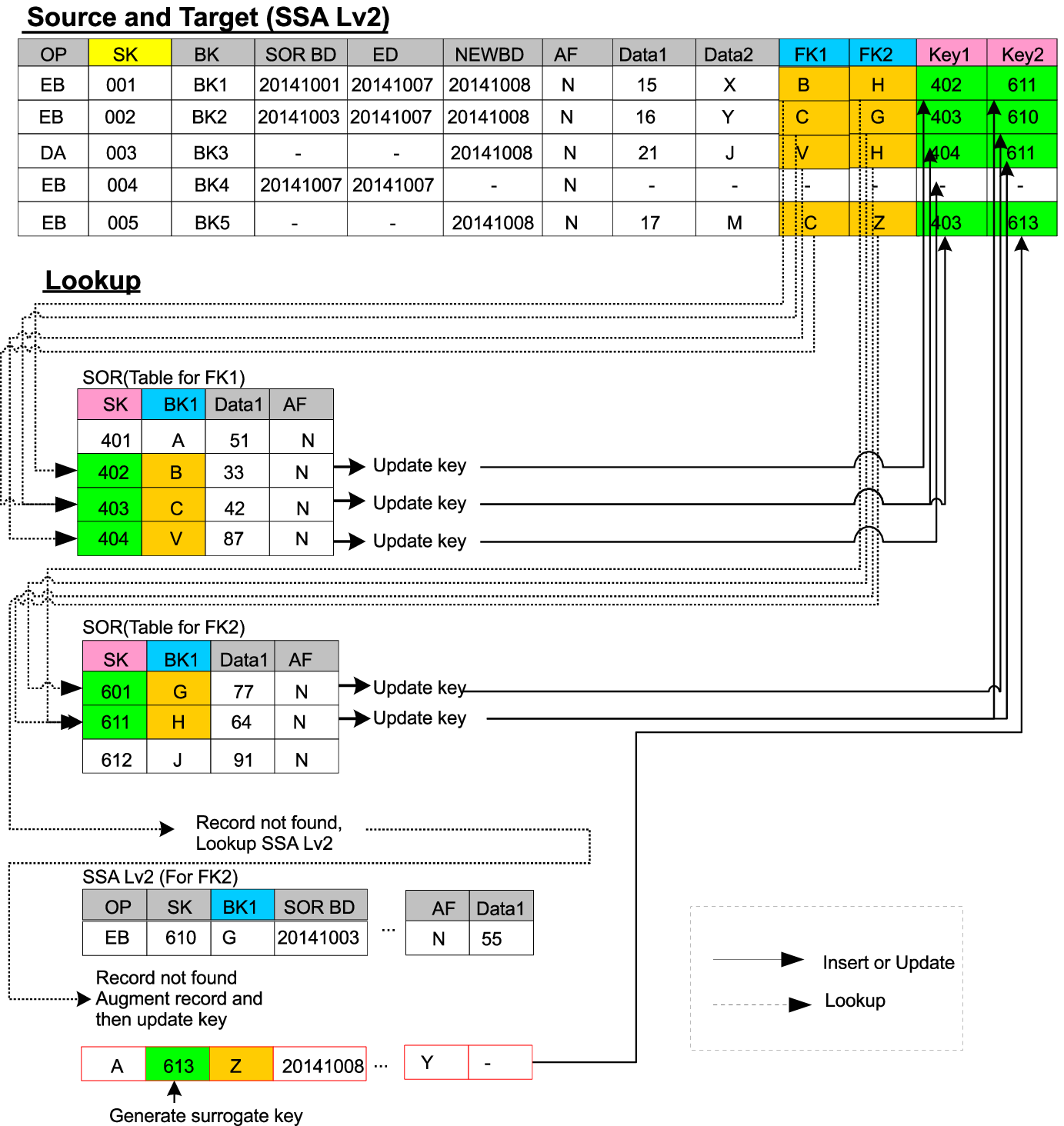, scale=0.8}
\vspace{-20pt}
\caption{Key validation}
\label{fig:keyvalidation}
\vspace{-15pt}
\end{figure*}

We now use a running example to illustrate the transformation process (see Figure~\ref{fig:transformation}). As shown, the process involves source table,  lookup table, and target table. The figure describes the generic logic of the change detection and how the records are moved from SSA Lv1 to SSA Lv2 base on the different operation codes. In this example, we consider the following four scenarios: 1) For the records streamed from SSA Lv1 with the business key values, BK=``BK1" and BK=``BK2", the corresponding records with the surrogate keys (SK=``001" and ``002") are found in SOR. Therefore, ``EB" operation is identified and added to the operation code field, {\em OP}, in the SSA Lv2 table. The surrogate keys are added to the field of SK. The transaction date (TxDate) is added to the field of new begin date (New BD), and the end date (ED) is set to one day ahead of TxDate (due to the daily ETL job). Other attributes in the SSA Lv2 table are mapped from the corresponding attributes of the SSA Lv1 table. 2) For the record with BK=``BK3" from the SSA Lv1 table, an argument record is found in the SOR table with SK=003, and its argument flag is on (AF=1).  The operation ``DA" is identified, and the attribute values from SSA Lv1 are added to SSA Lv2. The values will be used to update the empty attributes in the argument record in SOR. 3) For the record from SSA Lv1 with BK = ``BK4" and TxCode = ``D", we look up the SOR table using the business key value, and the record is found in SOR with the surrogate key value, ``0004". Therefore, an ``E" operation is identified for this record. The SOR record will be marked as ``delete" in the end by updating the end date. The other attribute values do not have to be mapped from the SSA Lv2 table for the ``delete" record. 4) For the record from SSA Lv1 with BK = ``BK5" and transaction type code ``I", we cannot find any previous version in the SOR table. The ``B" operation is identified, which represents the new record created in the source system. Therefore, the surrogate key is generated, and the last record transaction date in SSA Lv1 is updated to the field of new begin date (New BD) in the SSA Lv2 table.

\subsection{Key Validation}
\label{sec:keyvalidation}
In an SOR table, the foreign key values should be the surrogate key values from its reference tables, rather than the business key values as used in the source system. Recall that we mapped the foreign key values from the source system during the transformation process, but we have not updated the actual foreign key values in the SOR table.  We delay building the foreign key relationship to the key validation process. The benefit is that whenever the data of a table is ready, we can start the transformation job without having to wait for the foreign key reference tables, which allows us to process the early-arriving data in a timely manner. Algorithm~\ref{alg:keyvalidation} shows the key validation process. The process looks up the surrogate key value from the foreign-key referenced tables in both SOR and SSA Lv2 using business key values, and then uses the returned surrogate key value to update the foreign key field. The lookup was done on both SOR and SSA Lv2 tables. It is because the SOR table only stores the records up to previous ETL run, while the records of current run are still stored in the SSA Lv2 table. Before key validation starts, all the business key values that are required for the foreign key lookup are moved into the SSA Lv2 table by the transformation process.

When a foreign key cannot be found both in SOR and SSA Lv2 table, a blank record containing only the surrogate key and business key is created and added into the referenced table (see line~10--11). This process is called {\em augmentation}. The augmentation process enables the key validation job to be run in parallel without waiting for the transformation job on the foreign key reference table. It saves much time since we can first process the early-arriving data into child tables. The  augmentation ensures that the referential integrity constraint in SOR is not violated.  Since we do not have to consider the loading order of the tables with dependencies using the argumentation method,  it is possible to design parallel ETL workflow for a better performance.
\begin{algorithm}[htp]
\caption{Key Validation}
\label{alg:keyvalidation}
{\scriptsize
\begin{algorithmic}[1]
  \Require{Data has been loaded into SSA Lv2}
  \State $row \gets$ Read a record from SSA Lv2
  \State $found$ $\gets$ Lookup SOR reference table for key values
  \If{$found$}
  	\State Update $row$ with the found key values
  \Else
	\State $found \gets$ Look up SSA Lv2 reference table for key values
	\If{found}
		\State Update $row$ with the found key values
	\Else
		\State $sk \gets$ Get a new surrogate key
		\State Add an argument record with $sk$ and OP=A, and insert into SSA Lv2
	\EndIf
  	
  \EndIf
\end{algorithmic}
}
\end{algorithm}

Figure~\ref{fig:keyvalidation} shows the key validation process using the running example. Note that the target table for foreign key update is the  SSA Lv2 table. We detail the key validation process using the following two scenarios: 1) In the SSA Lv2 table, the records with surrogate key ``0001", ``0002", and ``0003" are streamed in. The FK1 and FK2 fields store the foreign key values extracted from the source table. The business key values in FK1 and FK2 are used to look up the SOR table which is the foreign-key referenced table. The lookup finds all the surrogate keys, and the keys are updated back to the fields of Key1 and Key2 in the SSA Lv2 table, respectively. 2) In the SSA Lv2 table, the record with the surrogate key ``0005" is streamed in. The business key values in FK1 and FK2 are used for looking up the SOR table. The business key ``C" is found in the SOR table for FK1, but ``Z" cannot be found for FK2. We, then, further look up the foreign-key referenced table for FK2, \ie,  the SSA Lv2 table. The lookup fails again. Since the lookup for ``Z" failed in both SOR and SSA Lv2 tables, we create an augmentation record (or dummy record) with the business key value, ``Z" and the  surrogate key value, ``613", and  insert the record into the SSA Lv2 table. We also turn on the augment flag to indicate that it is an argumentation record. But, we now have the surrogate key value ``613" that could be used to update the field, Key1, in the SSA Lv2 table.
\begin{table*}[htp]
\centering
\caption{SSA Lv2 table}
\begin{tabular}{|c|c|c|c|c|c|c|c|c|c|c|c|c|}
\hline
\rowcolor[HTML]{C0C0C0} 
\textbf{OP} & \textbf{SK} & \textbf{BK} & \textbf{SOR BD} & \textbf{ED} & \textbf{NEW BD} & \textbf{AF} & \textbf{Data 1} & \textbf{Data 2} & \textbf{FK1} & \textbf{FK2} & \textbf{Key1} & \textbf{Key2} \\ \hline
EB          & 001         & BK1         & 20141001        & 20141007    & 20141008        & N           & 15              & X               & B            & H            & 402           & 611           \\ \hline
EB          & 002         & BK2         & 20141003        & 20141007    & 20141008        & N           & 16              & Y               & C            & G            & 403           & 610           \\ \hline
DA          & 003         & BK3         & -               & -           & 20141008        & N           & 21              & J               & V            & H            & 404           & 611           \\ \hline
E           & 004         & BK4         & 20141006        & 20141007    & -               & N           & -               & -               & -            & -            & -             & -             \\ \hline
B           & 005         & BK5         & -               & -           & 20141008        & N           & 17              & M               & C            & Z            & 403           & 613           \\ \hline
\end{tabular}
\label{tab:ssalv2}
\end{table*}
\begin{figure*}[htp]
\begin{minipage}[b]{0.5\linewidth}
\centering
\epsfig{file=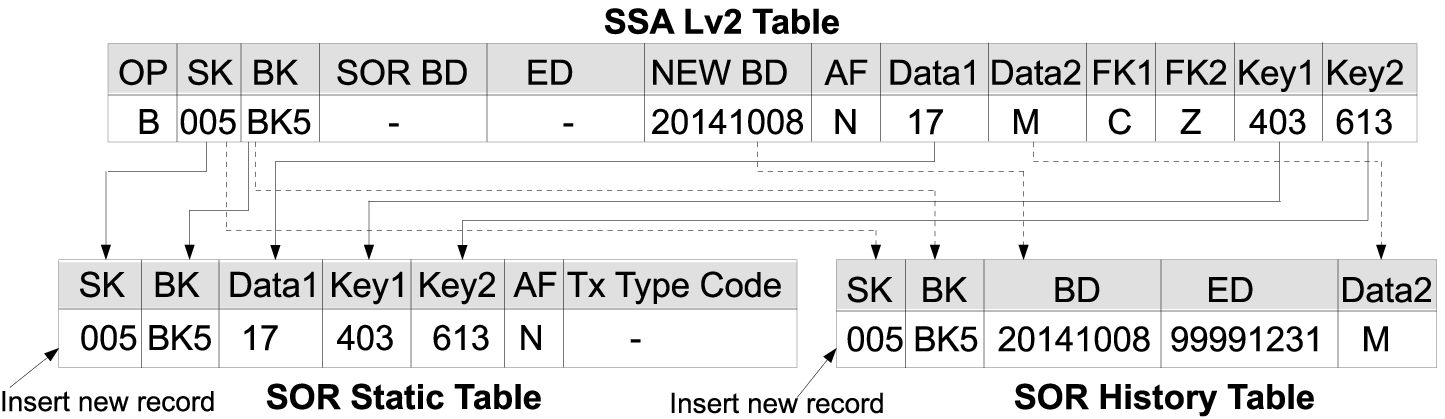, scale=0.62}
\vspace{-5pt}
\caption{Loading scenario~1}
\label{fig:load1}
\end{minipage}
\hspace{0pt}
\begin{minipage}[b]{0.5\linewidth}
\centering
\vspace{-5pt}
\epsfig{file=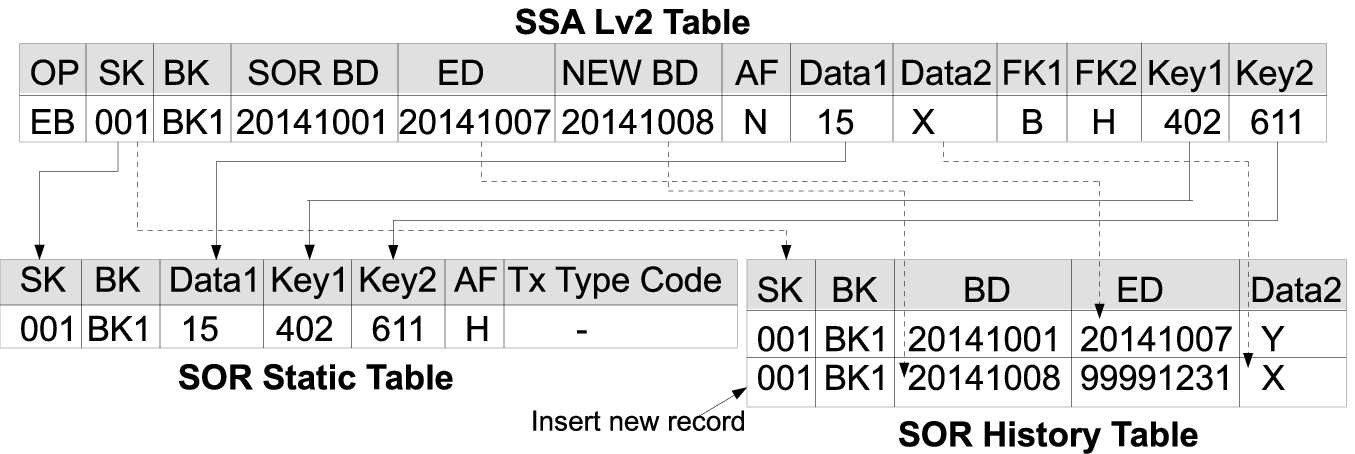, scale=0.62}
\vspace{-5pt}
\caption{Loading scenario~2}
\label{fig:load2}
\end{minipage}
\vspace{-5pt}
\end{figure*}

\begin{figure*}[htp]
\begin{minipage}[b]{0.5\linewidth}
\centering
\epsfig{file=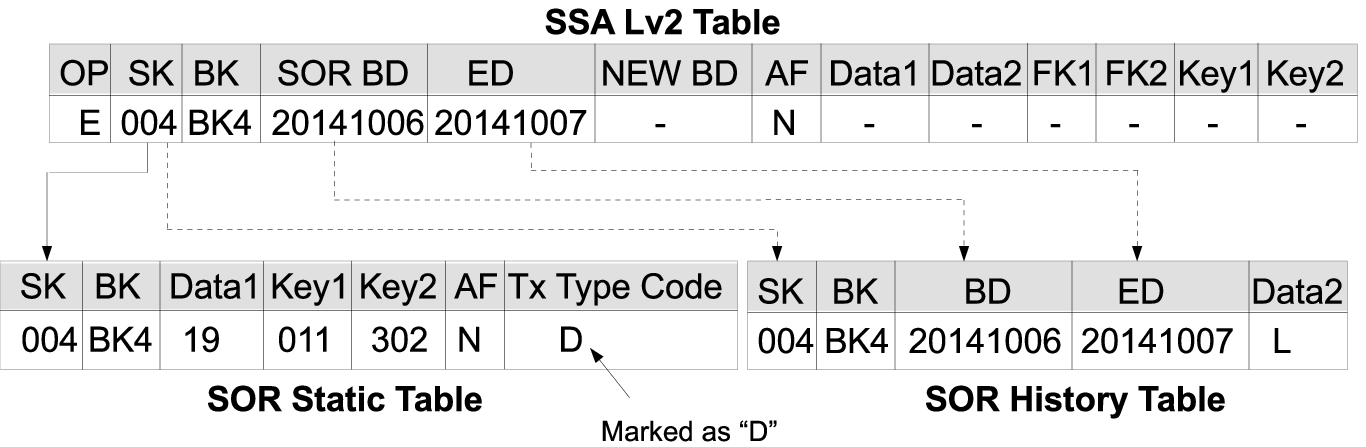, scale=0.62}
\vspace{-5pt}
\caption{Loading scenario~3}
\label{fig:load3}
\end{minipage}
\hspace{-10pt}
\begin{minipage}[b]{0.5\linewidth}
\centering
\epsfig{file=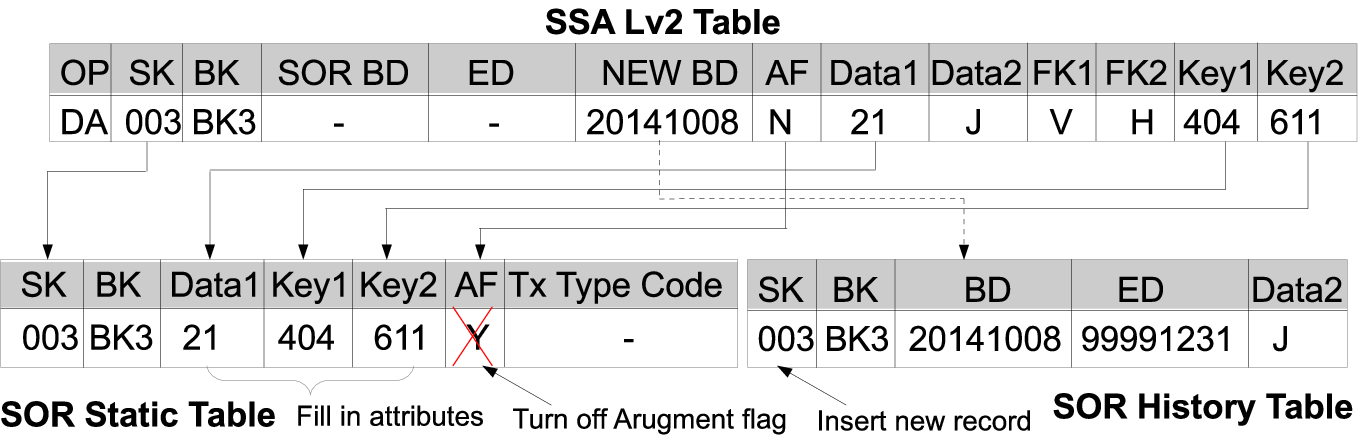, scale=0.62}
\vspace{-5pt}
\caption{Loading scenario~4}
\label{fig:load4}
\end{minipage}
\vspace{-5pt}
\end{figure*}

\subsection{Loading}
After the processes of transformation and key validation, the records in SSA Lv2 are ready to be loaded into SOR tables. We classify the data to be loaded into two  categories, static data and dynamic data. Static data refer to those unlikely change over time, such as customer personal information including name, birthday, sex, etc.,  while  dynamic data refer to those more likely to change, such as customer address, and phone, etc. In SOR, the static data are simply stored in the static table while the dynamic data are stored in the history table with begin date and end date for tracking the changes. During the transformation and key validation processes,  static and dynamic data both were moved to the same SSA Lv2 table, which will be moved to the corresponding static and history tables in SOR by the loading process. Recall that we have identified different operation codes during  transformation and key validation processes, and stored the codes in the SSA Lv2 table. In the loading process, we move the data into SOR tables based on these operation codes. The loading process is described in Algorithm~\ref{alg:loading}

After the data being processed by the transformation and key validation, we have the data in the SSA Lv2 table, which are ready for the loading (see Table~\ref{tab:ssalv2}). The loading process involves the source table in SSA Lv2, and the target tables in the SOR. Compared with the  transformation and key validation, the loading process is much simpler, which no lookup action is required. We identify the following five typical loading scenarios, which load  static and dynamic data into SOR tables based on the operation codes.

{\bf Scenario 1.} Figure~\ref{fig:load1} shows the loading process for the records with the operation code ``B".   For the static data from the SSA Lv2 table, we insert a new record with the surrogate key ``005", business key ``BK5" into  SOR static table. For the dynamic data from the SSA Lv2 table, we insert a new record into the history table with surrogate key ``005", business key ``BK5", begin date ``20041008", and end date ``99991231" (representing the latest record).

\begin{algorithm}[htp]
\caption{Loading Process}
\label{alg:loading}
{\scriptsize
\begin{algorithmic}[1]
  \Require{Data are ready in the SSA Lv2 table}
  \State $row \gets$ Read a record from the SSA Lv2 table
  
  \If{$row[OP]=B$}
  	\State Insert into static table
  	\State Insert into history table
  \ElsIf{$row[OP]=EB$}
  	\State Update  static table for the changed values.
    \State Update  history table, and set the end date. 
    \State Insert the record into history table: with same surrogate key as the previous version;  with new begin date;  with ``99991231" as the end date; and with the latest values for the other fields.
  \ElsIf{$row[OP]=D$}
  	\State Update static table, and set the last transaction type code as ``D";
  	\State Update history table, and set the end date.
  \ElsIf{$row[OP]=E$}
  	\State Update static table with the latest static values, and turn off the augment flag.
  	\State  Insert the record into history table: with same surrogate key as static table; with a new begin date;  with ``99991231" as the end date; and with the latest values for the other fields.
  \ElsIf{$row[OP]=DA$}	
  	 \State Update static table with the latest static values, and turn off the augment flag.
\State  Insert the record into history table: with same surrogate key as static table; with a new begin date; with ``9999-12-31" as the end date; and with the latest values for the other fields.
  \ElsIf{$row[OP]=A$}	
    \State Insert the record into static table: with a new surrogate key generated; with a new begin date; with ``9999-12-31" as the end date; and with the augment flag turned on.
  \EndIf
\end{algorithmic}
}
\end{algorithm}

{\bf Scenario 2.}
Figure~\ref{fig:load2} shows the loading process for the record with surrogate key ``001" and the operation code ``EB". Since the SOR static table has not kept any history record, we can directly insert the static data into the SOR static table. For the SOR history table, we have to first look up  the record using the surrogate key ``001" and SOR begin date ``20141001". When the record is found, we first update the end date of the history record to ``20041007". Then, we insert into a new record with the identical surrogate key ``001" and business key ``BK1",  fill the begin date ``20141008" and the dynamic data from the SSA Lv2 table, and set the end date to ``99991231", which represents the latest record.

{\bf Scenario 3.}
Figure~\ref{fig:load3} shows the loading process for the record with surrogate key ``004" and operation code ``D". We first locate the record in the SOR static table using the surrogate key, then set the last record transaction type code as ``D", representing that the record has been deleted.  In the SOR history table, we locate the record using the surrogate key ``004" and SOR begin date ``20141006". If the record is found, we update the end date ``20141007" that is from the SSA Lv2 table.

{\bf Scenario 4.}
Figure~\ref{fig:load4} shows the loading process  for the record with surrogate key ``003" and operation code ``DA". In the SOR static table, we locate the augment record  using the surrogate key. Since the augment record only contains surrogate key, business key, and augment flag, we update the blank attributes, and turn off the augment flag. In the SOR history table, we insert a new record with surrogate key ``003", business key ``BK3", begin date ``20141008", and the dynamic data from the SSA Lv2 table. The end date is set to ``99991231" to represent the latest record.

\begin{figure}[htp]
\centering
\epsfig{file=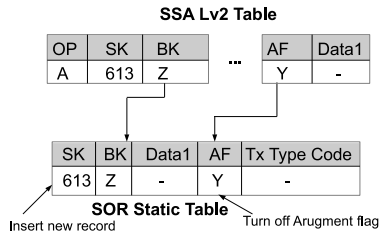, scale=0.65}
\vspace{-5pt}
\caption{Loading scenario~5}
\label{fig:load5}
\vspace{-10pt}
\end{figure}

{\bf Scenario 5.}
Figure~\ref{fig:load5} shows the loading process  for the record with surrogate key ``613" and operation code ``A". In the SOR static table, we insert a new record with surrogate key ``613", business key ``Z", and augment flag ``1". The other fields are left blank. Since it is for augmentation, we do not need to insert any record into the SOR history table. The argument record will be updated again when the record with ``DA" from SSA Lv2 arrives next time.

\section{DDS Data Process}
According to the ETL process shown in Figure~\ref{fig:architecture}, it is the DDS data process after the two-level data staging ETL process. DDS data process summarizes and aggregates the data from SOR, and inserts into the dimension and fact tables in DDS, or directly transforms the data to multi-dimensional OLAP cubes (MOLAP). The DDS data process does dimension and fact extraction. The dimension extraction process extracts dimension records from SOR based on the last transaction date, and only extracts the records whose last transaction dates are equal to or greater than the starting date of the current ETL job. If a record in SOR with the last transaction type code ``D", the record will not be extracted to DDS if the target is a static dimension table; But, if the target is a slowly changing dimension (SCD) table, the record will be extracted to keep track of changes. Compared with the dimension extraction, fact extraction is more complicated. We have to determine which record to be extracted based on the actual affected date and the last transaction date. To identify which field in SOR is the actual affected date, we need to examine what information the fact table is going to store. For example, for the sales data mart, if the fact table is for holding order data, the actual affected date should be the transaction date in the SOR transaction order table. After identifying the actual affected date, we select all the records in the SOR table with the actual affected date,  and process them into the corresponding DDS fact table. If the fact table contains the measures calculated from the measures of the previous date, all the records in the SOR table at or after the earliest affected date need to be retrieved to rebuild the fact table.

\section{Related Work}
\label{sec:relatedwork}
ETL has been the subject of extensive research in the academic community. The recent papers \cite{vassiliadis2009,thomsen2009} present a survey of the research in ETL.  The ETL design has been discussed from different perspectives. The papers \cite{trujillo2008,luja2005}  use UML-based method to design  ETL processes, and \cite{Tryfona,Sapia} use ER model in the ETL design.  Akkaoui and E.~Zim\'{a}nyi \cite{akkaoui} propose a platform-independent ETL conceptual model based on the Business Process Model Notation (BPMN) standard, and implement ETL using Business Process Execution Language (BPEL). Skotas and Smitsis \cite{skotas,alkis2003} introduce the semantic web technologies,  {\em ontology}, to the  ETL conceptual modeling. In \cite{alberto2012,oscar2012}, the author presents the requirement-driven approach for the ETL design, where the ETL can be produced separately for each requirement, and  consolidated incrementally. Thomsen and Pederson propose a Python-based ETL programing framework to address ETL programming efficiency, \eg, only a few code lines are needed to implement an ETL program for a specific dimensional schema \cite{pygrametl2009}. In this paper, we, on the other hand, use the bottom-up approach for the ETL design. The proposed ETL method aims at simplifying processing transaction data, and provides ``one-stop" approach for handling early-arriving, fast- and slowly-changing data.

Optimizing ETL is a time-consuming process, but it is essential to ensure ETL jobs to complete within  specific time frames. For ETL optimization,  Simitsis \et propose a theoretical framework \cite{simitsis2005,simitsis20051,Simitsisqos10}, which formalizes ETL state spaces into a directed acyclic graph (DAG), then searches the best execution plan with regards to the time of the state spaces. Tziovara \et propose the approach of optimizing ETL based on an input logical ETL template \cite{Tziovara20051}. In \cite{LZ05}, Li and Zhan analyze the task dependencies in an ETL workflow, and optimize ETL workflow by applying the parallelization technique to the tasks without dependency. Behrend and J\"{o}rg use rule-based approach to optimize ETL flows \cite{BJ10}, where the rules are generated based on the algebraic equivalences. Our approach is an ETL method, but can also be used to optimize the ETL in solving the early-arriving data problem, in which no dependencies need to be considered.

The latest trend of data warehousing is to support big data, and offer real-time/right-time capability, \eg, \cite{alfredo2014,rite}. The emergence of the cloud computing technologies, such as MapReduce \cite{mapreduce}, makes it feasible for ETL to process large-scale data on many nodes. As the evidence, the two open source MapReduce-based systems, Pig \cite{Olston2008} and Hive \cite{hive2009}, become increasingly used in data warehousing. But, they both are designed for the generic purpose for big data analytics, with limited ETL capabilities, somewhat like  DBMSs other than full-fledged ETL tools. To complement this, our previous work,  ETLMR \cite{etlmr,etlmr2012}, extends the ETL programming framework \cite{thomsen2009,thomsen2011} using MapReduce, but maintains its simplicity in implementing a parallel dimensional ETL program. Besides, the framework \cite{cloudetl2014} is proposed to improve the dimensional ETL capability of Hive. The difference is that ETLMR aims at the traditional RDBMS-based data warehousing system, while the latter uses Hive for more scalability.

\section{Conclusion}
\label{sec:conclusion}
In data warehousing, it is challenging to process transaction data featured with high frequent changes. In this paper, we have proposed a two-level data staging ETL method for handing the transaction data. The proposed method identifies the changes of source data into different operation codes, then pre-processes the data from the staging area of the level-one to the level-two, and finally loads the data into a data warehouse based on the operation codes. In the paper, we used a running example to illustrate how to use the proposed ETL method, including detecting the changes, identifying the operation codes, verifying the key, and loading. The proposed ETL provides the ``one-stop" approach for processing fast-changing, slowly-changing, and early-arriving data.


\end{document}